\documentclass[twoside,fleqn]{article}
\usepackage{espcrc2,epsf,amssymb}

\title{Dynamical Properties of Large $N$ Reduced Model of Yang-Mills
  Theory}

\author{T. Hotta\address{Institute of Physics, University of Tokyo,
    Komaba, Meguro-ku, Tokyo 153-8902, Japan}
  ,
  J. Nishimura\address{Department of Physics, Nagoya University,
    Chikusa-ku, Nagoya 464-8602, Japan}
  and
  A. Tsuchiya\address{Department of Physics, Graduate School of
    Science, Osaka University, Toyonaka, Osaka 560-0043, Japan}}

\begin{document}

\begin{abstract}
  We study the large $N$ reduced model of $D$-dimensional Yang-Mills
  theory with special attention to the dynamical aspects related to the
  eigenvalues of the $N \times N$ matrices, which correspond to the
  space-time coordinates in the IIB matrix model.
  We define a quantity which represents the uncertainty of the
  space-time coordinates and show that it is of the same order as the
  extent of the space time, which means that the classical space-time
  picture is maximally broken.
  The absence of the SSB of the Lorentz invariance is also shown.
\end{abstract}

\maketitle

\section{Introduction}
Recently the large $N$ reduced models \cite{EK} have been revived in
the context of nonperturbative formulations of string theory
\cite{BFSS,IKKT}.
The IIB matrix model \cite{IKKT,FKKT,AIKKT}, which is the reduced
model of ten-dimensional supersymmetric Yang-Mills theory, is expected
to be a nonperturbative formulation of superstring theory.

In this article, we study the large $N$ reduced model of Yang-Mills 
theory, which we refer to as the ``bosonic model'' in what follows.
It is nothing but the bosonic part of the IIB matrix model.
The action is given by
\begin{equation}
  S= -\frac{1}{4 g^2} \mbox{Tr} ([A_\mu,A_\nu]^2),
\end{equation}
where $A_\mu$ $(\mu = 1, \cdots, D)$ are $N \times N$ traceless
hermitian matrices.

We investigate the dynamical aspects of the model related to the
eigenvalues of $A_\mu$, which correspond to the space-time
coordinates in the IIB matrix model.
One of the most important quantities is the extent of the space time
defined by
$R=\sqrt{\langle \frac{1}{N} \mbox{Tr} (A^{2}) \rangle}$. 
We also define a quantity which represents the uncertainty of the
space-time coordinates and an order parameter for the SSB of the
Lorentz invariance.
We determine these values for the bosonic model by Monte Carlo
simulation.

\section{The extent of the space time}
\label{extentsp}

For $D$ $>$ 2 with sufficiently large $N$, the integral of $A_\mu$ is
convergent without any cutoff because of an attractive potential
between the eigenvalues of $A_\mu$ \cite{KS}.
This enables us to absorb $g$ by rescaling $A'_\mu = A_\mu / \sqrt{g}$
, which means that $g$ is nothing but a scale parameter.
The large $N$ behavior of $R$ is, therefore, parametrized as $\sqrt{g}
N^{\omega}$, since $R$ is proportional to $\sqrt{g}$ on dimensional
grounds.
In the IIB matrix model, $\omega$ plays an important role to deduce
the space-time dimension, which is to be determined dynamically from
the eigenvalue distibution of $A_\mu$ \cite{NT}.
We determine $\omega$ for the bosonic model.

The observables we measure are the following.
\begin{eqnarray}
  (a)& &
  \left \langle \frac{1}{N} \mbox{Tr} (A^2) \right\rangle \times 
  \left( \sqrt{g}N^{1/4} \right)^{-2}, \nonumber \\
  (b)& &
  \left\langle \frac{1}{N} \mbox{Tr} ((A_\mu A_\nu ) ^2 )
  \right\rangle \times \left( \sqrt{g}N^{1/4} \right)^{-4}, \nonumber
  \\
  (c)& & 
  \left\langle \frac{1}{N} \mbox{Tr} ((A^2)^2) \right\rangle \times 
  \left( \sqrt{g}N^{1/4} \right)^{-4}, \nonumber \\
  (d)& &
  \left\langle \frac{1}{N} \mbox{Tr} (F ^2) \right\rangle \times 
  \left( \sqrt{g}N^{1/4} \right)^{-4},
\end{eqnarray}
where $F_{\mu\nu}= i[A_\mu, A_\nu]$.
Here, we have normalized the above quantities so that they are
dimensionless, and hence they are independent of $g$.
We note that $(d)$ can be obtained analytically as $(d) = D (1 -
1/N^2)$ \cite{HNT}.

\begin{figure}[htbp]
  \begin{center}
    \leavevmode
    \epsfysize=0.39\textwidth
    \epsffile{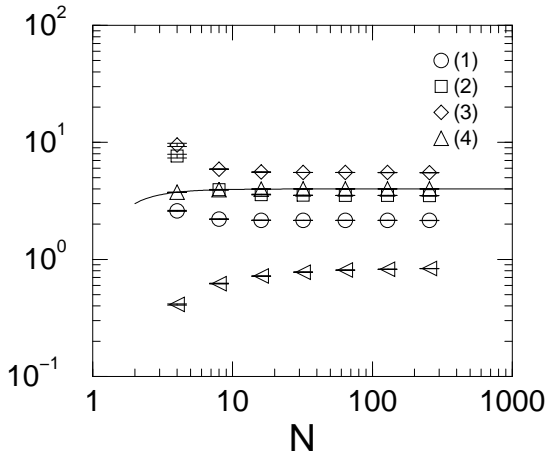}
    \caption{$(a)$ $\sim$ $(d)$ are plotted for $D=4$.
      The solid line represents the exact result for $(d)$.
      The tilted triangles are the plot of $\left\langle \Delta ^2
      \right \rangle / (\sqrt{g}N^{1/4})^2$, which is discussed in
      Section \ref{bd_classics}.}
    \label{fig:D4ulog}
  \end{center}
\end{figure}

\begin{figure}[htbp]
  \begin{center}
    \leavevmode
    \epsfysize=0.39\textwidth
    \epsffile{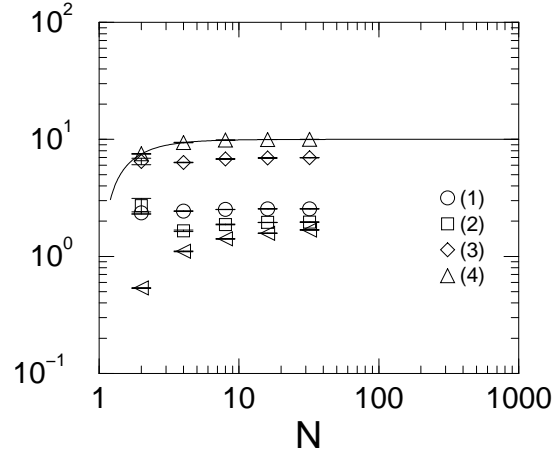}
    \caption{$(a)$ $\sim$ $(d)$ are plotted for $D=10$.
      The solid line represents the exact result for $(d)$.
      The tilted triangles are the plot of $\left\langle \Delta ^2
      \right \rangle / (\sqrt{g}N^{1/4})^2$, which is discussed in
      Section \ref{bd_classics}.}
    \label{fig:D10ulog}
  \end{center}
\end{figure}

In the Fig. \ref{fig:D4ulog}, we show our results of Monte Carlo
simulation for $D$ $=$ 4 with $N$ $=$ 4, 8, 16, 32, 64, 128, 256.
One can see that $(a)$ $\sim$ $(d)$ are constant for $N \gtrsim 16$.
In Fig. \ref{fig:D10ulog}, we plot the data for $D$ $=$ 10 with $N$
$=$ 2, 4, 8, 16, 32.
Here again we find that $(a)$ $\sim$ $(d)$ are constant for $N \gtrsim
16$.

From these results, we find that $R$ is of the order of
$O(\sqrt{g}N^{1/4})$, which means that the upper bound determined by
the perturbative argument is saturated.
The large $N$ behaviors of $(a) \sim (c)$ are consistent with the
predictions from the perturbative expansion and the $1/D$ expansion of
this model \cite{HNT}.

\section{Breakdown of the classical space-time picture}
\label{bd_classics}

In the IIB matrix model, the eigenvalues of $A_{\mu}$ represent the
space-time coordinates.
If $A_{\mu}$'s are commutable, they can be diagonalized simultaneously
and the diagonal elements can be regarded as the
classical space-time coordinates.
Therefore it makes sense to ask to what extent the classical
space-time picture is broken.
To discuss this issue, we define a quantity which represents the
uncertainty of the space-time coordinates.

We define such a quantity by the analogy to the quantum mechanics.
We regard the matrices $A_\mu$'s as the linear operators which
act on the linear space, which we identify as the space of states
of particles.
We take an orthonormal basis $|e_i \rangle$ ($i=1,2,\cdots,N$) and
identify $|e_i \rangle$ with the state of the $i$-th particle.
The space-time coordinate of the $i$-th particle is defined by
$\langle e_i | A_\mu | e_i \rangle $.
The uncertainty of the coordinate can be defined by
\begin{equation}
  \delta (i)^2
  = \sum_{\mu} \left\{
    \langle e_i | A_\mu ^{~2} | e_i \rangle
    - (  \langle e_i | A_\mu | e_i \rangle )^2  \right\}.
\end{equation}
We take an average of $\delta (i)^2$ over all the particles and define
the gauge-invariant quantity by
\begin{equation}
  \Delta^2 = \min_{\{|e_i \rangle\} } \left( \frac{1}{N} \sum_{i}
    \delta(i)^2 \right). 
\end{equation}
Note that $\Delta = 0$ if and only if $A_\mu$ are diagonalizable
simultaneously.

We plot $\langle \Delta^2 \rangle$ obtained by Monte Carlo simulation
in Figs. \ref{fig:D4ulog} and \ref{fig:D10ulog}.
We find that $\langle \Delta ^2 \rangle / (\sqrt{g}N^{1/4})^2$ tends
to a constant for large $N$.
Thus we conclude that $\sqrt{\langle \Delta ^2 \rangle}$ is of the
same order as $R$.
If $\Delta$ were smaller than the typical distance between two
particles, i.e., $\ell = R N^{-1/D}$, we might say that the classical
space-time picture is good.
Our result, however, shows that the classical space-time picture
is maximally broken.

\section{No SSB of Lorentz invariance}

The SSB of Lorentz invariance is of paramount importance in the IIB
matrix model, since if the space time is to be four-dimensional, the
10D Lorentz invariance of the model must be spontaneously broken.

The SSB of Lorentz invariance can be probed by
\begin{equation}
  J = \frac{1}{D} I_{\mu\nu} I_{\mu\nu} - \left( \frac{1}{D}
    I_{\mu\mu} \right)^2 ,
\end{equation}
where $I_{\mu\nu}$ is defined by $I_{\mu\nu} = \mbox{Tr} (A_\mu
A_\nu) / N$.
$J$ represents nothing but the variation of the eigenvalues of
$I_{\mu\nu}$.

If $\langle J \rangle/ R^4 $ is nonzero in the large $N$ limit, the
Lorentz invariance is spontaneously broken.
Thus, $\langle J \rangle / (\sqrt{g}N^{1/4})^4 $ can be considered as
an order parameter of the SSB of Lorentz invariance.
In Ref.\cite{HNT}, we show that when the large $N$ factorization 
holds, the order parameter goes to zero in the $N\rightarrow \infty$
limit.
Indeed, we show that at all orders of the $1/D$ expansion, the
factorization holds and the order parameter behaves as $1/N^2$.

\begin{figure}[htbp]
  \begin{center}
    \leavevmode
    \epsfysize=0.39\textwidth
    \epsffile{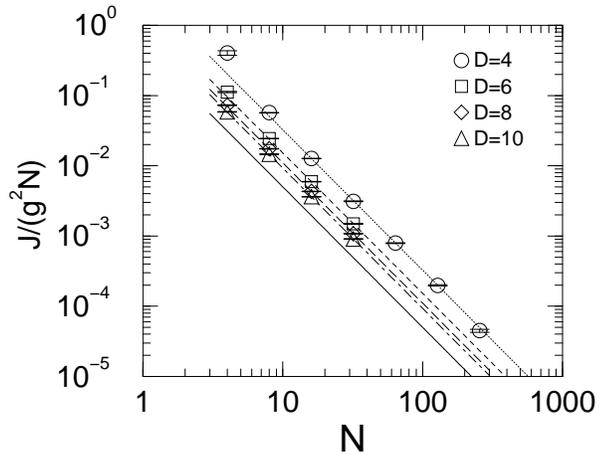}
    \caption{$\langle J \rangle/ (\sqrt{g}N^{1/4})^4$ is plotted
      against $N$ for $D$ $=$ 4, 6, 8,10.
      The solid line represents the prediction for the $D \rightarrow
      \infty$ limit and the other lines represent the fit to the
      $1/N^2$ behavior \cite{HNT}.}
    \label{fig:5D}
  \end{center}
\end{figure}

In Fig. \ref{fig:5D} we plot $\langle J \rangle / (\sqrt{g}N^{1/4})^4$
against $N$ for $D$ $=$ 4, 6, 8, 10.
We see that the order parameter vanishes in the large $N$ limit,
which means that the Lorentz invariance is not spontaneously broken.
Note also that the large $N$ behavior of the order parameter can be
nicely fitted to $1/N^2$.
This is in accordance with the prediction by the $1/D$ expansion.

\section{Summary}

We studied the dynamical aspects of the reduced model of bosonic
Yang-Mills theory.
We determined the extent of the space time and showed that the
classical space-time picture is maximally broken.
We also showed that Lorentz invariance is not spontaneously broken.
We expect that our findings for the bosonic model provide a helpful
comparison when we investigate the IIB matrix model.

\end{document}